# The biased interaction game: Its dynamics and application in modelling social systems.

Phil Mercy <p.j.n.mercy@qmul.ac.uk>

Martin Neil <m.neil@qmul.ac.uk>

Queen Mary University of London

10th December 2025

Version 0.47

## Abstract

The biased interaction game described the operation of systems rooted in boundedly rational interactions under conditions of scarcity. The game explored the influence of bias and demonstrated how hierarchy and inequality are emergent system properties when sources of bias, such as power and scarcity, affect the outcome of interactions in an environment. Bias also impacts the likelihood of the emergence of cooperation.

This paper serves as a companion piece to the paper introducing the biased interaction game. It investigates the general applicability of the game and demonstrates how the consideration of bias can modify and improve upon prior systems thinking. In particular, it shows how social systems can be successfully modelled using the biased interaction game and confirms its suitability for modelling extreme examples such as hyper-capitalism and social egalitarianism. It also reveals how biased systems can demonstrate non-linear behaviour, where long periods of system stability are punctuated by short bursts of rapid hierarchical transitions, mimicking real-world observations of social mobility.

The paper concludes with a simplified real-world application, modelling the merits of two competing wealth redistribution philosophies: 'social welfare' and a 'universal basic income'

# 1. Introduction

The biased interaction game is a novel game-theoretic treatment of agent interaction. The game introduced two new factors: scarcity in the environment, and an agent's incumbent value, and it was fully described and detailed in a prior publication (Mercy & Neil, 2025). The approach was built upon an assumption of boundedly rational rules of interaction, where scarcity and incumbent value are both able to influence the outcome of any interaction.

Incumbent value, or the accumulated wealth, power, or influence of an individual agent, was introduced to facilitate a bias on the outcome of any interaction between unequal agents. Consideration of scarcity in an environment provides a mechanism for the scaling of the impact of bias, with incumbent value being of greater importance in scarcer environments.

Systems modelled by the biased interaction game are self-organising in character, and hierarchy emerges naturally from a minimal set of rules. The approach thus lends itself to the examination of the origins of organisational structure of a number of systems, including human social systems.

The approach resulted in simulated systems that exhibited the natural emergence of hierarchy and inequality, both aspects deepening with increasing scarcity. The likelihood of cooperation between agents was also subject to bias: whilst cooperation in the same hierarchical band was always favoured, cooperation between different bands in the hierarchy became increasingly less likely as scarcity increased.

This paper further explores the dynamics of these systems, and their applicability, but avoids an exhaustive statistical analysis of the model. Instead, the intent is to illustrate the general characteristics of system models based on the biased interaction game to enable a comparison with other approaches. The paper documents a series of open-loop experiments conducted to reveal the inherent nature of systems modelled using the game, and game variables are modified one by one to reveal their impact upon emergent system behaviour. Complex concepts such as feedback, and evolution, are deliberately excluded from this exercise to better allow the principal characteristics and behaviours to reveal themselves.

The paper begins with a brief overview of the defining characteristics of the biased interaction game approach. It then explores three edge-case examples analogous to utopian equality, hyper-capitalism, and social egalitarianism, both to discover how the game behaves when confronted with extreme but plausible scenarios, and to better understand the limits of its applicability. The paper moves on to explore the concept of social mobility within an emergent hierarchy, and the underlying movement of individuals between bands of a hierarchy whose overall structure is resilient to such movement.

Finally, a simple biased interaction game model is constructed to simulate a test of two competing philosophies of social wealth redistribution: social welfare, and the universal basic income. The exercise reveals how new insight can be gained using this novel method and illustrates how the approach might benefit other real-world scenarios too.

# 2. Introduction to the biased interaction game

A brief summary of the biased interaction game as introduced in Mercy & Neil (2025) will be beneficial at this juncture. The origins of the biased interaction game lie in the system-theoretic observation that most interactions between individual agents seem to be biased in some way, favouring one agent over another. This effect, widely described in economics literature as 'market power' (De Loecker et al., 2020), influences many systems wherever a mechanism for bias exists.

A second observation, building on the work of Yudkowsky (2017) and Christensen (2016) amongst others, is that agents tend to act in a boundedly rational manner, maximising their own benefit given the information available to them, regardless of any system-wide considerations.

These observations together led to a game design that incorporated two primary factors: an agent's incumbent value (analogous to accumulated power, influence, or wealth) and an environment's scarcity (the availability of, or access to, the source of utility or value within an environment).

In the biased interaction game, we consider a system as comprising of a number of agents who interact, in an environment that shapes the outcome of those interactions. We initially consider the interaction of two agents. Agents have a choice of two strategies: 'cultivate' (active engagement with the environment) or 'utilise' (passive reliance on the environment). The benefits of each strategy are dependent both on the scarcity in the environment, and the relative incumbent values of the two interacting agents. Agents are assumed to choose an optimal mix of the two strategies when interacting with another agent, to maximise the benefit to them from that interaction.

Each interaction results in the distribution of a pot of value. The pot itself consists of contributions from each of the two interacting agents, the size of which is governed by their relative incumbent values when compared to the incumbent value of all other agents in the environment. The pot is distributed at the conclusion of an interaction in proportions determined by the payout equations of the biased interaction game. These equations relate the scale of the payout to both the relative incumbent values of the agents interacting and the scarcity in the environment.

Whether one agent or the other does better out of each interaction depends on the proportion of the pot awarded to them when compared to their initial contribution to the pot, reaching an equilibrium where the award matches the pot contribution exactly.

Conceptually, an environment is biased both by its scarcity and by the difference in incumbent values of the agents operating within it. In environments without scarcity, it is natural to be passive, and agents will allow the environment to provide for them. As scarcity increases there is a greater impetus for agents to take an active role in the environment, to facilitate provision from the limited utility available.

One can conceive of scenarios where it is boundedly rational for agents to adopt passive-passive, passive-active, or active-active relationships with other agents they encounter. This allows for a rich model of varied interaction topologies whose nature has the capacity to be influenced by bias, both in the environment and between the agents interacting within it.

It is most informative to review the payout equation for one particular interaction example, a passive-active interaction. Here, two agents $X$ and $Y$ interact, with $X$ adopting a utilise (passive) strategy and $Y$ a cultivate (active) strategy. Equation (1) shows the share of the pot awarded to agent $X$, $Share_X$, where $S$ ($0 \leq S \leq 1$) is the level of scarcity in the environment, $M_X$ is the incumbent value of agent $X$, and $M_Y$ is the incumbent value of agent $Y$.

$$Share_X = \frac{(M_Y(1-S))}{(M_X+M_Y)} \quad (1)$$

In general, if agent $X$ has a lower incumbent value than agent $Y$, then it does well by adopting the utilise strategy, as this strategy favours the weaker of the two agents interacting. More formally, in an environment with low scarcity ($S \to 0$), it is boundedly rational for agent $X$ to adopt a utilise strategy when its incumbent value is lower than that of agent $Y$.

As scarcity $S$ increases, however, the advantage for an agent with relatively weak incumbent value in adopting the utilise strategy erodes, prompting a switch to a more active strategy. In practice, a point of equilibrium is reached in any environment where the effect due to the difference in agent incumbent value is offset by the biasing effect that scarcity has on interactions in the environment.

In an environment with higher scarcity ($S \rightarrow 1$), it becomes increasingly more beneficial for an agent to adopt a cultivate strategy, even for agents with inferior incumbent value. Agents with superior incumbent value always do better by adopting a cultivate strategy.

Whenever scarcity is non-zero ($S \neq 0$), a point of equilibrium exists for a given ratio of incumbent values, where neither agent gains nor loses from the interaction, despite one having a superior incumbent value. The incumbent values of agents interacting iteratively will tend to grow or shrink until this ratio of incumbent values is reached between them.

For a system consisting of a population of interacting agents, environments with little or no scarcity are likely to disadvantage agents with initially superior incumbent value. This would lead over the course of successive interactions to a redistribution of incumbent value from the strongest to the weakest agents, and a decrease in inequality between them. There would be no incentive to adopt an active strategy, so a system with low levels of scarcity is characterised by weakly interacting, passive, agents.

An increase in scarcity creates an environment where it is more beneficial to have superior incumbent value. The balancing effect between scarcity and incumbent value superiority, however, results in a tendency for systems to self-organise after successive interactions into hierarchical bands of agents, with constant incumbent value ratios between agents in different bands. This is akin to social strata in human social systems.

Agent-based modelling was used in Mercy & Neil (2025) to investigate system level behaviour as agents in the biased interaction game repeatedly interacted. The experiments revealed emergent inequality in the population as agents stratify into hierarchical bands of like wealth or power. It found a conservation of cooperative behaviour between agents within a single hierarchical band, but a degradation in the tendency for agents from different bands to cooperate as scarcity increased. Whilst this result mirrored social hierarchy in human society and demonstrated theoretically how the emergence of cooperative behaviour can be influenced by scarcity and inequality, the paper acknowledged that further work was required to fully understand the applicability of this modelling approach.

## 3. Operation at the limit – edge-case scenarios

To better understand the possibilities and limitations of the biased interaction game approach, a number of extreme scenarios (edge-cases) were investigated through experimentation. Each experiment was conducted using agent-based modelling of a biased interaction game environment, written in Python, using the Mesa ABM library (Kazil et al., 2020).

Each experiment is defined by a choice of variables as follows:

- Scarcity $S$ ($0 \leq S \leq 1$) defines the degree of scarcity in the environment. A value of $S = 0.3$ has been shown to produce hierarchies with Gini coefficients broadly similar to the wealth distributions of countries such as the US and UK.
- The number of agents $N$ is chosen to be as high as possible (to minimise the effect of atypical starting distributions for agent incumbent value). Typically, in this series of experiments, $N = 50$.

- The interaction risk ratio $R$ ($0 \leq R \leq 1$) scales the proportion of each agent's incumbent value that is at risk in any single interaction in the model. This defines the amount an agent contributes to the interaction pot that is redistributed at the conclusion of every interaction. In practice, high values such as $R = 0.95$ can lead to environments where changes in the fortune of an agent can happen rapidly, and emergent results can appear quickly. Lower levels lead to slower dynamics that can be analysed with higher granularity at the conclusion of an experiment, and some aspects of agent dynamics may be less likely to occur at all.
- The length of an experiment is measured in terms of the number of iterations performed, $I$. An iteration of the model is a calculation phase where each agent interacts at least once with another randomly chosen agent in the environment.
- An experiment begins with each agent having an initial incumbent value. The scheme for defining the distribution of starting values is a part of the definition for each experiment. These starting values are critical to the outcome. Some of the following experiments use fixed starting points, such as the experiment looking at an egalitarian initial distribution (referred to as 'Equal'). Others require a more representative starting distribution, such as a randomly generated one. For traceability and reproducibility, a randomly generated starting distribution was generated once, for a population of 50 agents, and then reused in multiple experiments (referred to as 'RAND50').

A formal measure of the level of inequality arising in the environment in each experiment is provided by the Gini coefficient for the distribution of the agent incumbent values. The Gini coefficient is an established and widely used measure of wealth inequality. Whilst not universally seen as an appropriate measure in all instances (Kuijt, 2024), the coefficient has been widely used in the social sciences and economics to explore the impact of inequality (Vona & Patriarca, 2011).

A scatterplot chart is used to visualise the outcome of each experiment, where each point represents an agent in the environment. The x-axis identifies the incumbent value an agent has at the start of an experiment and the y-axis its incumbent value at an experiment's conclusion. Any bias due to initial incumbent value, or any self-organisation effect emerging, is readily apparent in the chart. This type of chart is used in Figures 1, 6, 7, and 8.

Three edge-case scenarios are considered analogous to a utopian equality, hyper-capitalism, and social egalitarianism. The results are plotted in Figure 1's charts a, b, and c respectively.

3.1 All agents created equal (utopian equality)

An experiment (3.1) is designed to discover whether hierarchies still emerge if all agents start a simulation with the same incumbent value. This is an important scenario to explore to determine whether egalitarianism is an inherently stable state in a biased interaction game environment.

Experiment 3.1:

- $S = 0.3$
- $N = 50$
- $R = 0.95$
- $I = 2000$
- Starting distribution = 'Equal' ($= \frac{1}{N}$)

The result can be seen in Figure 1a. All agents start and finish with the same incumbent value. The resulting distribution remains egalitarian with a Gini coefficient = 0.00. This experiment demonstrates that a lack of differential incumbent value amongst agents at the start of the experiment leads to zero diversity at its completion, implying that a truly egalitarian system operating under biased interaction game principles is inherently stable, even when there is scarcity in the environment.

## 3.2 All agents cultivate (hyper-capitalism)

Experiment 3.2:

- $S = 0.3$
- $N = 50$
- $R = 0.95$
- $I = 2000$
- Starting distribution = 'RAND50'

Experiment 3.2 is designed to discover the implications of the universal adoption of the cultivate strategy. One could conceive of practical mechanisms that might severely limit the adoption of one strategy or another (religious, ideological, mechanical, or technical) and it is important to understand the consequences of such a limitation at the system level.

The resulting distribution is shown in Figure 1b, and the agent population at the conclusion of the experiment had a Gini coefficient of 1.0. The distribution of incumbent value is greatly impacted when all agents cultivate. Most agents finish with near zero incumbent value, whilst the remaining environment value is shared amongst a small number of winners (two in this experiment).

One might imagine that the experiment should end with only one winner, with other participants having virtually zero incumbent value. In fact, the precise number of winners will be highly dependent on the initial starting distribution of incumbent value, and random chance in early interactions for agents with relatively high starting values will determine whether they prosper or not. If they interact only with agents with lesser incumbent value in early encounters, then they could build up sufficient incumbent value to survive their first interaction with an agent with superior incumbent value. In this way hierarchical bands might still emerge but with only one or two members in all but the lowest band. The lowest band would contain the vast majority of agents (and have a vanishingly small average incumbent value).

## 3.3 All agents utilise (social egalitarianism)

Experiment 3.3:

- $S = 0.3$
- $N = 50$
- $R = 0.95$
- $I = 2000$
- Starting distribution = 'RAND50'

Experiment 3.3 is designed to discover the implications of the universal adoption of the utilise strategy. The opposite of the 'always cultivate' scenario is one where all agents adopt a passive (utilise) stance in all interactions.

When agents only adopt a utilise strategy, the distribution of incumbent value amongst them collapses, as can be seen in Figure 1c. All agents end the experiment with equal incumbent value, regardless of their starting value, and the Gini coefficient for the resulting distribution is 0.0.

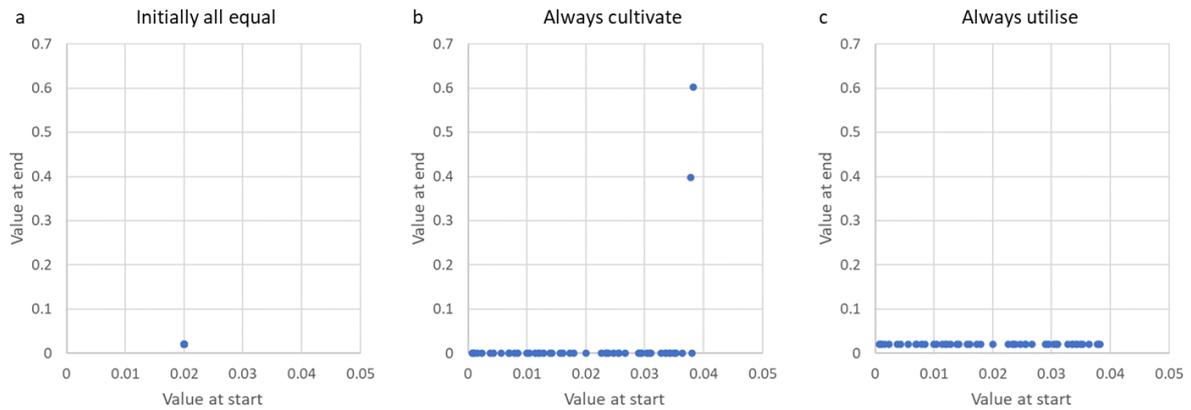

*Figure 1 – Start vs end incumbent value for edge-case experiments 3.1 (a), 3.2 (b), and 3.3(c)*

3.4 Edge-case scenario conclusions

These results suggest that, whilst absolute equality is inherently stable, cultivation tends to increase inequality whilst utilisation decreases it. Operating without constrained strategies or starting positions, it is likely that systems will stabilise where the effect of these two strategies balance each other out. The precise mix of the strategies adopted, and the agents adopting them, will depend on the starting distribution of agent incumbent values and the degree of scarcity in the environment. One can deduce that with scarcity comes a degree of inequality that is inherently stable.

# 4. Social mobility

Experiment 4 is designed to examine the movement of agents over time and, in particular, to determine whether their hierarchical destiny is fixed and dependent upon their starting incumbent value. The experiment is tailored to reveal if there is evidence for social mobility in a system modelled using the biased interaction game, in the form of movement between hierarchical bands.

Experiment 4:

- $S = 0.3$
- $N = 50$
- $R = 0.95$
- $I = 50,000$
- Starting distribution = 'RAND50'

A high value for $R$ is chosen to provoke social mobility in the experiment as the scope for an agent's incumbent value to change rapidly is increased. With large values of $R$, a high proportion of an agent's incumbent value is reserved to finance all of its interactions with other agents. This represents a high risk for every agent and a series of favourable (or unfavourable) interactions can lead to large increases (or decreases) in their incumbent value.

The model is run for 50,000 iterations in total but is programmed to store the status of the agents every 100 iterations. This results in 500 chronological snapshots of the environment, and 500 incumbent value data points for each agent. For each snapshot, both the hierarchical banding and band membership for every agent are calculated.

The results are visualised using a colour-coded grid of cells showing a map of band membership. In this visualisation, the experimental results form a grid of 500 columns (snapshots) by 50 rows (agents). Each column is a snapshot in time of the status of each of the 50 agents in the experiment, with each cell colour coded to indicate their band membership. The execution timeline of the experiment is plotted from left to right, with the first snapshot being the leftmost column, and the rightmost column the last. Four bands are depicted from the lowest to highest average incumbent values, respectively: 'Band 1' (blue), 'Band 2' (orange), 'Band 3' (grey), and 'Band 4' (yellow).

Each row forms a timeline recording the changing band membership experience for an individual agent. When inspecting a row, any change in colour indicates that an agent has moved between bands in the hierarchy at some point in the preceding 100 iterations of the model, relative to its peer agents. Changing band membership is evidence for social mobility within the context of the experiment. The coloured grid for this experiment is shown in Figure 2.

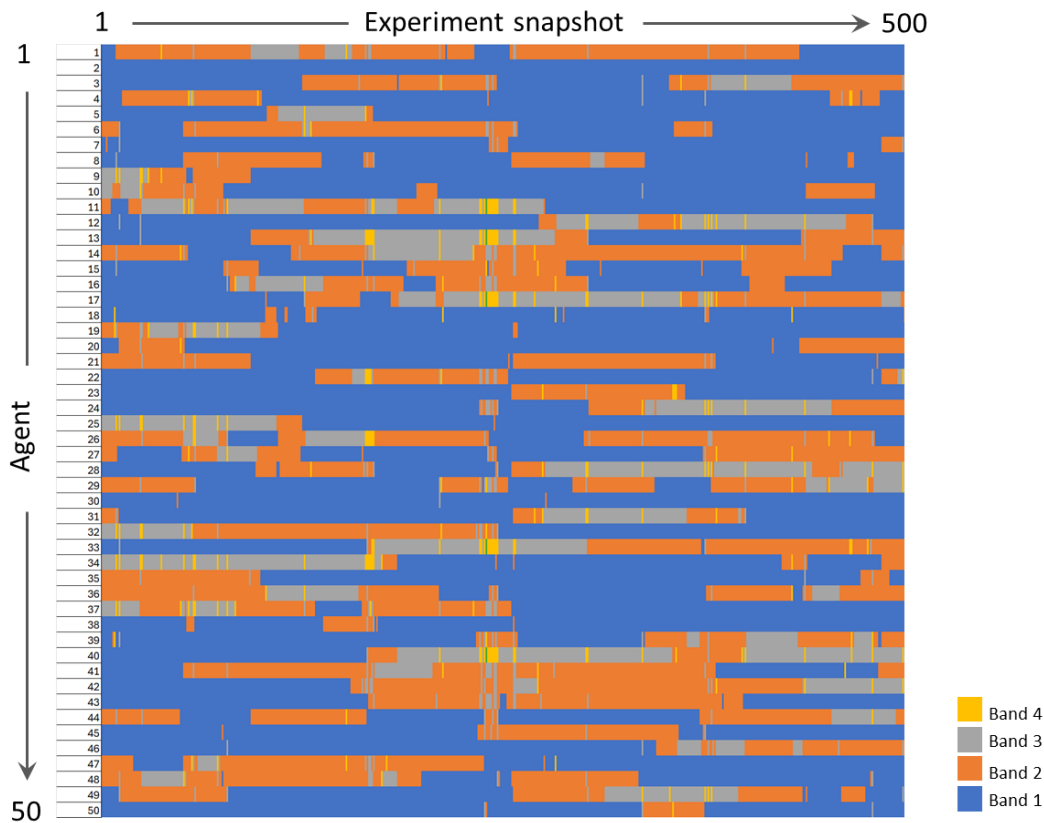

*Figure 2 – Band membership colour map for 50 agents over 50,000 model iterations (Exp 4)*

Figure 2 reveals how most agents transition between bands in the hierarchy multiple times under these experiment conditions. On average agents transitioned 23.2 times in the course of the 50,000 iterations. The conclusion is that social mobility is not only possible but potentially a frequent occurrence for agents in a system based on the biased interaction game, even though the underlying hierarchical structure that emerges remains intact.

A closer look at the results in Figure 2 reveals that agent band transitions don't appear to be evenly spread throughout the execution of the experiment. If we define an experimental era as a series of consecutive snapshots, then some eras appear to be relatively free of transitions, whereas other eras see a high proportion of agents transitioning (all within a relatively small number of iterations). Those eras with particularly high rates of transition are highlighted in Figure 3.

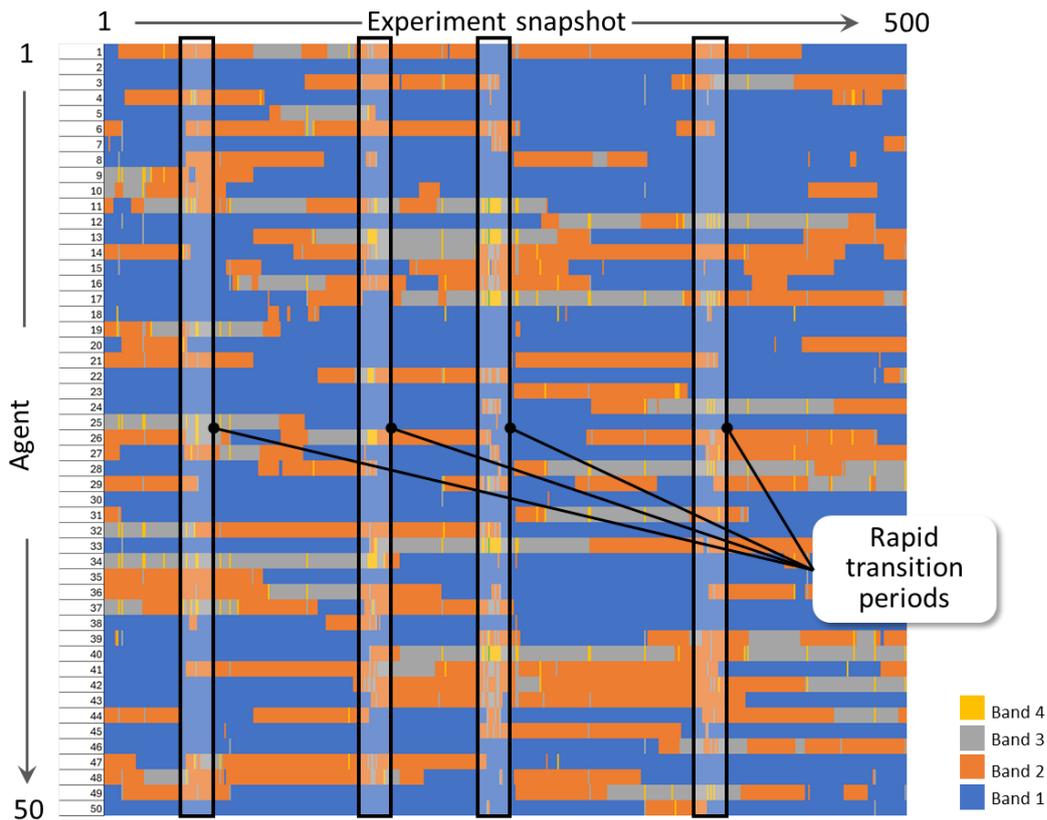

*Figure 3 – Colour map from Experiment 4 with eras of high transition rates highlighted*

This pattern of experiencing long periods of relative stability with occasional periods of rapid transition is a characteristic reminiscent of the rapid phase change often found in non-linear or chaotic systems. Complex non-linear systems commentator and entrepreneur Kieran Kelly has written at length about the defining characteristics of chaotic systems and in particular about complexity emerging from simple underlying mechanisms. Kelly would refer to this type of apparent randomness in operation as 'abrupt non-continuous change' (Kelly, 2019) and it would be reasonable to interpret this result as being indicative of a non-linear system.

## 5. Investigating non-linearity

Given the results presented in Section 4, a natural question to ask is whether social mobility in a system based on an iterated biased interaction game is a result of inherent non-linearity? To investigate non-linearity, a longer-term experiment is required.

Experiment 5:

- $S = 0.3$
- $N = 50$
- $R = 0.75$
- $I = 200,000$
- Starting distribution = 'RAND50'

Compared to the social mobility experiment in section 4, a lower risk value ($R = 0.75$) reduces the likelihood of agents transitioning between bands, but an increase in the duration ($I = 200,000$) provides the opportunity to detect the nature of such transitions if they do occur. The experiment records a snapshot of agent status every 20 iterations for a total of 10,000 chronological snapshots from the experiment. In particular, it was recorded whether an agent had transitioned from one band to another in the preceding 20 iterations.

The results are depicted in Figure 4, a chart that plots the experiment snapshot (x-axis) against the proportion of the agent population that transition in that snapshot, i.e. the 'transition density' (y-axis).

The results show long eras (successive snapshots) where few agents transition between bands of the hierarchy, but these are punctuated by eras where a large proportion of the agent population transitions.

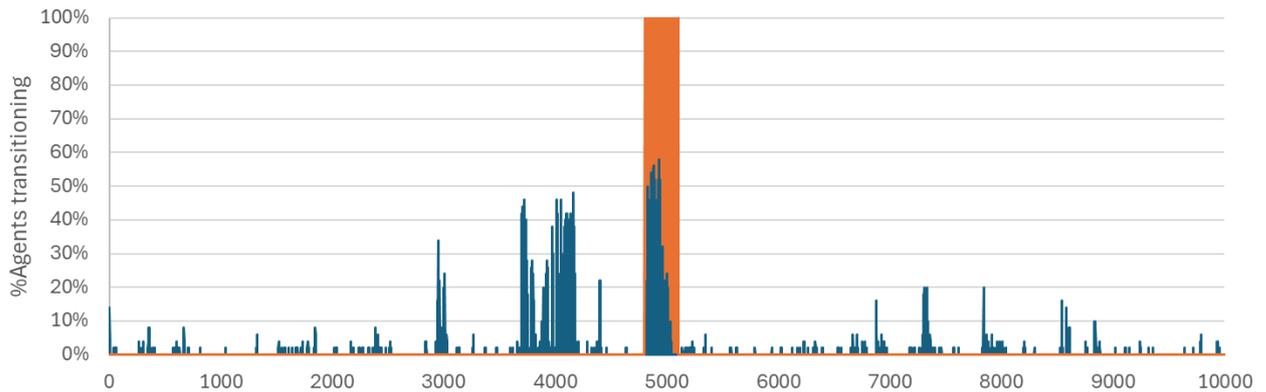

*Figure 4 – Long-term transition density over 10000 snapshots of 20 iterations each (Exp 5)*

One particular sequence of snapshots (4800 to 5100) is highlighted in orange in Figure 4 and zoomed into in Figure 5 to reveal more detail.

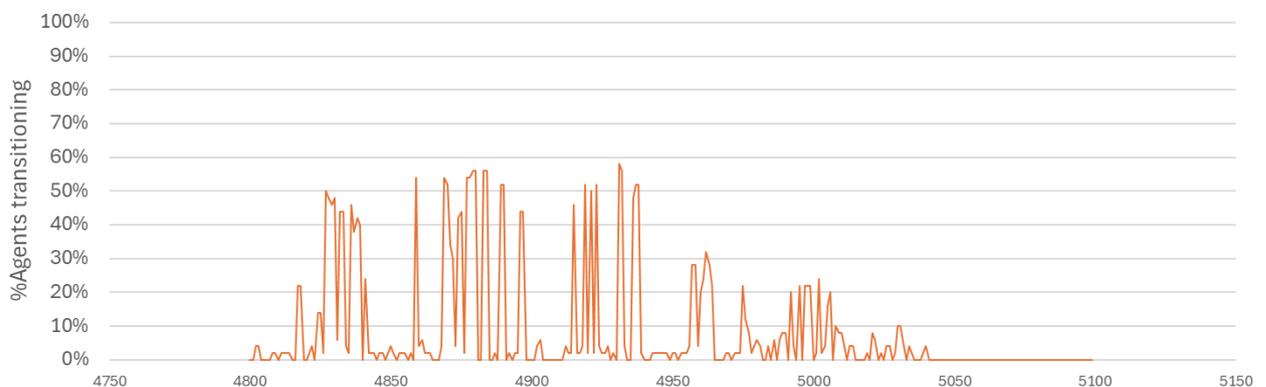

*Figure 5 – Transition density between snapshots 4800 and 5100 from Figure 4*

Once a group of agents with similar incumbent values have formed a band in the hierarchy, it might seem natural for each agent to remain a member of the same group indefinitely. However, these results reveal that, whilst agents usually remain in place, agents can and do freely transition between bands. Indeed, when such movement does occur, it can also trigger band transitioning in other agents, and it is possible for a high proportion of agents to transition in the course of relatively few interactions. Even so, the overall hierarchical structure that emerges is both robust and repeatable. The mechanism that allows widespread band transitioning within a hierarchical structure that remains resilient to such movement merits closer consideration.

Whilst a rare occurrence, several extremely favourable (or unfavourable) interactions occurring together in a short timespan may displace an agent away from its local hierarchical band, its incumbent value becoming significantly different to the band's mean.

Other agents interacting with such an agent are also likely to receive atypically favourable (or unfavourable) outcomes, since its incumbent value is so atypical for the environment, differing from the norm for any of the bands that have emerged. This is likely to result in larger than typical losses (or gains), shifting them away from their band's average incumbent value too. As more agents interact with an increasing number of locally shifted agents, these too will be shifted away from their local band average values in a rapidly escalating fashion.

If an agent's shift in incumbent value is sufficient, it will have an incumbent value closer to that of an adjacent band than its previous band in the hierarchy. In this circumstance, a succession of more typical interactions would lead to its incumbent value getting ever closer to this adjacent band's average incumbent value. To put it another way, the agent is effectively attracted to and captured by an adjacent band in the hierarchy and will have transitioned to it.

The dynamics of agent transition shown in Figures 4 and 5 reveal behaviour typical of a non-linear or chaotic system. If a state of escalating agent transitions occurs, the system experiences what can be thought of as a 'mobility cascade', initially triggered by a rare sequence of interactions for one agent.

The extent of the non-linear behaviour, or the propensity towards mobility cascade, is likely to be dependent on the level of interaction risk in an environment. Smaller values of $R$ would require an increasingly rare series of beneficial (or detrimental) interactions in a row for one agent to trigger a mobility cascade. Increasing the number of agents $N$ is likely to increase the probability of a mobility cascade and exacerbate the inherent non-linearity of the system. With higher $N$, the likelihood that any single agent would experience the critical series of interactions needed to create a significant shift in incumbent value increases.

It has previously been demonstrated that a system based on the biased interaction game has an emergent hierarchical structure (Mercy & Neil, 2025). Agents stratify into bands of like incumbent value, with a repeatable incumbent value ratio between adjacent bands, the size of which is dependent on the scarcity in the environment. These characteristics also remain true at scale, as shown in the Appendix. Bands with a lower average incumbent value have a higher number of members, whilst those with a higher average incumbent value have a lower number of members. The precise number of layers that emerge for a given number of agents depends on the distribution of their initial incumbent values, but the general pattern of banding that emerges is both robust and repeatable.

Yet, despite its self-organising nature and the hierarchical structure that such a system adopts, the fate of the agents within a system is far from pre-ordained (or dependent on their initial incumbent value). Whilst the overall banding structure is robust, individual agents are still able to transition between bands if the right circumstances exist. Indeed, longer term experiments (such as Experiments 4 and 5) reveal periods of rapid transition, a mobility cascade, where a large proportion of agents transition between the bands in a short period of time. It transpires that, whilst the system's structure is both robust and repeatable, its inherent non-linear nature provides a mechanism for social mobility within it.

# 6. Example use case: wealth redistribution

The self-organising properties of the biased interaction game make it a good candidate for exploring inequality in human social systems, particularly as they pertain to wealth redistribution policy.

Two competing wealth redistribution policies are explored using a biased interaction game model. The first policy is based on a social welfare model, the second on a universal basic income.

Modern social welfare has its roots in the social insurance programs of Otto Von Bismarck in the late 19th century German Empire (Frohman, 2008) and evidence suggests that Western countries spending at least a fifth of their GDP on a welfare state have significantly reduced levels of poverty (Kenworthy, 1999). Whilst the specific focus for welfare support, (old age, disability, sickness, unemployment, etc.) varies from country to country, social welfare schemes exist under authoritarian and democratic regimes alike (Mares & Carnes, 2009).

The first scheme modelled is analogous to a simplified social welfare model, where all members of a population are taxed to provide a pot of funds which is then redistributed equally amongst the poorest members of society. Social welfare is then compared to a second scheme based on the idea of a universal basic income.

A universal basic income (UBI) is often viewed as a radical and controversial approach to wealth redistribution. It espouses the award of a cash grant to all members of a society, regardless of their current circumstances or needs. The population are taxed to provide the cash needed for such an award, the tax contribution from each individual being proportional to their current wealth. Wealthier members of society contribute the most to fund the scheme, but all members receive the same payout.

The crux of the controversy lies in the UBI concept of awarding the same financial grant to members of the society who don't need it, as well as to those who do. Detractors suggest that this is a waste of resources, taking them away from where they might do the most good for society. UBI advocates cite the simple mechanics and reduction in bureaucracy, as well as its ability to reduce inequality in a way that inherently protects justice and liberty (Bidadanure, 2019), in defence of the approach. Whilst counterintuitive as an approach for many, real world experiments with UBI tend to report positive impacts on inequality.

A simple taxation model was used in modelling both social welfare and UBI. A percentage of each agent's incumbent value (the tax rate $T_e$) was removed at the conclusion of each iteration of the model and placed into a pot ready for redistribution amongst the agents. The act of taxation is considered to be a function of the environment, operating independently from the agents operating within it. For simplicity, a fixed tax rate is used rather than a progressive one that might increase the tax rate for agents with higher incumbent value (wealth).

Defining for a population of $N$ agents, with an environment tax rate of $T_e$, where $M_x$ is the incumbent value of agent x, the pot of cash collected for wealth redistribution $\Phi$ is:

$$\Phi = \sum_{x=1}^{N} T_e M_x \quad (2)$$

By taxing agents once per model iteration, this method resembles a total wealth-based taxation approach. This is possibly closer in mindset to a salary-based tax scheme, rather than a more complex tax model focused on individual agent interactions which would be more analogous to a purchase-based tax scheme (such as a value-added or sales tax). This simplified tax approach facilitates a more general comparison of the impact of the two schemes, rather than a definitive examination of any specific tax or wealth redistribution policy.

An agent's share of the redistributed tax pot $\Phi$ depends on its hierarchical band and is different for the two redistribution schemes. Consider a system with *n* hierarchical bands, where *band 1* has the lowest average incumbent value, and *band n* the highest. If the total number of agents in the i<sup>th</sup> band, *band i*, is $N_i$, the total number of agents in the system, $N$, is given by:

$$N = \sum_{i=1}^{n} N_i \tag{3}$$

Similarly, if the wealth redistributed to agents in *band i* is $\Phi_i$, the total amount of wealth redistributed $\Phi$ is given by:

$$\Phi = \sum_{i=1}^{n} \Phi_i \tag{4}$$

Experiment 6:

- $S = 0.3$
- $N = 50$
- $R = 0.5$
- $I = 5,000$
- Starting distribution = 'RAND50'
- $T_e$ is varied in each experiment

Tax was collected and redistributed at the conclusion of every iteration.

In the social welfare-oriented model, the tax pot was divided evenly between members of *band 1*, the lowest hierarchical band, only.

$$\phi_i = \frac{\phi}{N_i} \tag{5}$$

$$where\ i = 1\ (otherwise\ 0)$$

In the UBI model, the tax pot was distributed with equal shares for every member of the population.

$$\phi_i = \frac{\phi}{N} \tag{6}$$

The results are visualised using a scatterplot chart of agent incumbent value with axes: x-axis, starting value and y-axis, ending value.

Figure 6 illustrates the resulting agent incumbent value distributions for a social welfare-oriented model, at different levels of taxation. To aid interpretation, the leftmost chart in Figure 6 depicts a baseline result with 0% tax, and hence no wealth redistribution occurring at all.

With 0% tax, the agents arrange themselves into three distinct hierarchical bands. The typically hierarchical nature of incumbent value distribution starts to break down at even small levels of taxation for the social welfare-oriented model. A 5% tax level leads to a preserved upper hierarchical band with slightly less incumbent value on average than the highest band seen without taxation. There is, however, a total blurring of distinction between the lower and the middle bands.

Increasing levels of tax lead to a collapse of the upper band. At 50% tax, there is a re-emergence of the upper band and a total collapse of middle and lower bands into a single lower band. Whilst a social welfare model reduces apparent inequality (and a Gini coefficient that decreases as tax increases supports this) this result suggests that higher tax rates can be extremely destructive to the natural hierarchy in the system and, in the extreme, cause a collapse of middle bands to form a populous underclass with a small and dominant upper oligarchy.

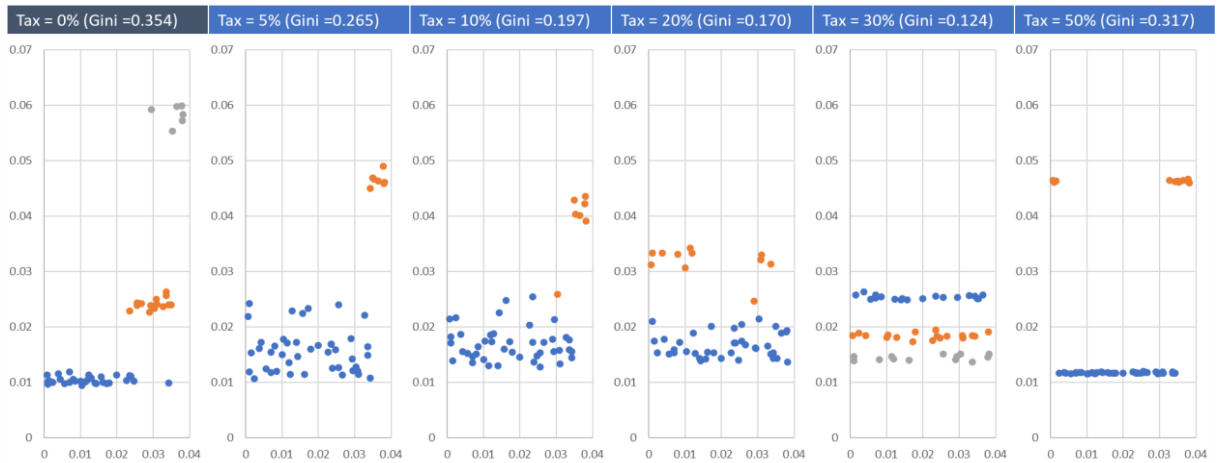

*Figure 6 – Agent incumbent value distribution in a social welfare model as tax varies (Exp 6)*

A social welfare approach employing lower rates of tax is shown in Figure 7. Here it is apparent that even small levels of tax (1%) are enough to disrupt the distinction between middle and lower hierarchical bands. Whilst social welfare may increase the incumbent value of many members of the lowest band, the impact is offset and borne mostly by members of the middle band, many of whom are adversely impacted.

One key observation is that, with the blurring of the distinction between middle and lowest bands, one's ability to accurately determine an agent's membership of the lowest band (and hence their eligibility for a social welfare payment) is made increasingly difficult. As a result, whilst technically capable of reducing inequality, a social welfare-based redistribution approach is likely to lead to slow and bureaucratic processes unable to assess eligibility as individual circumstances rapidly change.

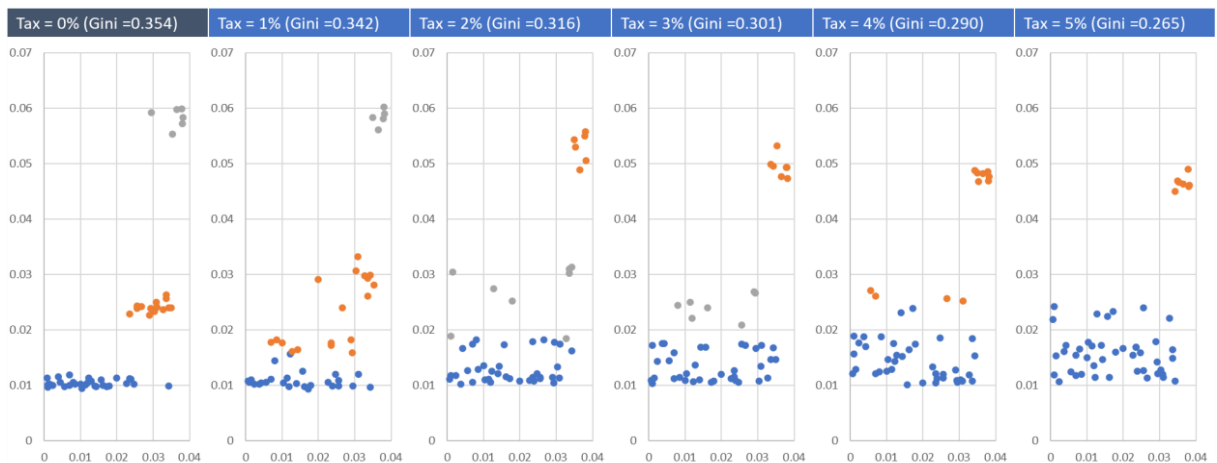

*Figure 7 – Agent incumbent value distribution at lower tax levels, for social welfare (Exp 6)*

The population incumbent value distributions for a UBI based approach are shown in Figure 8.

As tax rate increases, the distance between successive bands in the social hierarchy reduces, effectively reducing inequality (as evidenced by a reduction in the Gini coefficient). Whilst the average incumbent value of the highest band reduces, that of the lowest band increases. The reduction in inequality is achieved without destroying the inherent hierarchical structure of the society. Hence, for moderate levels of tax, the net result of implementing a universal basic income is to create an environment where the apparent (perceived or experienced) level of scarcity is reduced, whilst the overall hierarchical structure is retained (the wealthiest members of society are still the wealthiest).

Where the tax rate is higher, the hierarchy does begin to break down, with fewer bands and a greatly reduced distance between them. At a tax rate of 50%, the hierarchy collapses completely, and the result is an entirely egalitarian society with an equal incumbent value for all agents.

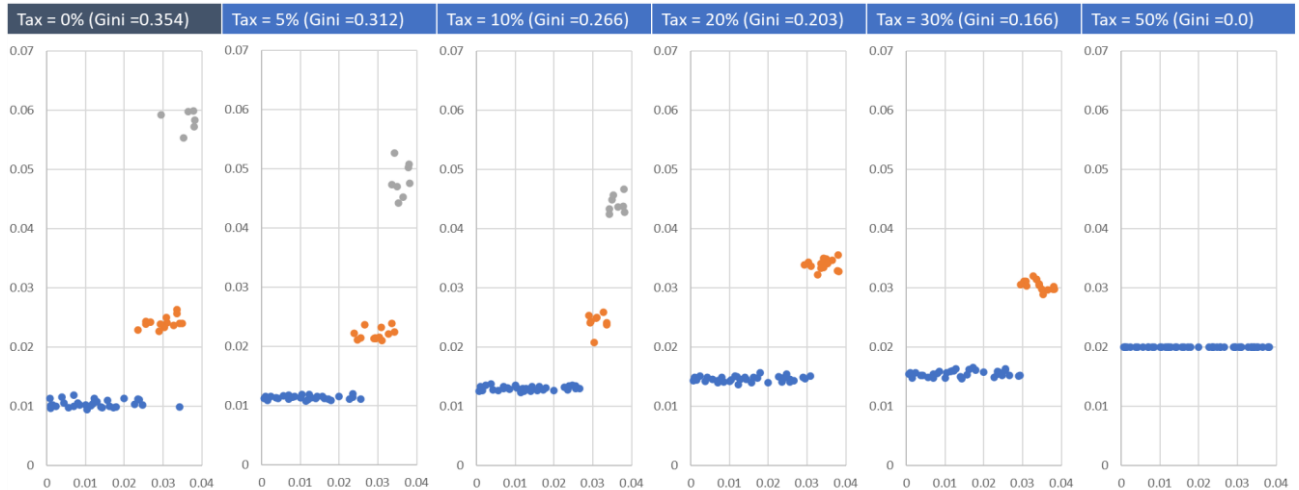

*Figure 8 – Agent incumbent value distribution as tax rate varies for a universal basic income (Exp 6)*

As expected, and as critics would certainly suggest, implementing a UBI requires a higher rate of tax to drive a reduction in inequality. However, the UBI approach retains the hierarchical social structure for moderate levels of tax in stark contrast to the social welfare-based redistribution model.

One could say that the big difference between the two approaches is that the UBI reduces inequality whilst maintaining stability, whereas a social welfare approach reduces inequality with a lower tax burden, but at the expense of the natural social hierarchical structure. The social welfare approach creates volatility in the lowest bands of society and in practice will likely engender an increased burden of administration with its inherent unfairness and bureaucracy.

## 7. Conclusion and discussion

This paper documents an initial exploration of biased interaction game-based models and their practical use.

Section 3 explored the limits of the model with extreme or edge-case scenarios. If all biased interaction game agents are created equal, with the same incumbent value at the start of an experiment, then they will remain equal no matter how many interactions between them occur. This is true even if there is scarcity in the environment. If game agents are constrained such that they always adopt a utilise strategy, then any differences between them are gradually eroded, and this results in an egalitarian system of total equality amongst the agents. If, however, agents always cultivate, then the vast majority of agents end up with a vanishingly small incumbent value, whilst the total value in the environment is shared between a small number of winners. This resembles an oligarchy.

The biased interaction game is capable, therefore, of producing environments with emergent structures anywhere between total equality and maximal inequality.

We observe that: in a social egalitarianism scenario where agents only utilise, equality is emergent, regardless of the level of scarcity in the environment. To provide a real-world illustration, such an environment might broadly resemble a society run under Marxist principles. In a hyper-capitalism scenario where agents only cultivate, inequality is emergent, irrespective of an environment's level of scarcity. In turn, a real-world analogue for this might be a version of an anarcho-capitalist society. Looking at both these extreme systems through a biased interaction game lens reveals something significant: neither system is able to react to (or correct for) any scarcity in the environment. This might be a valid cause for criticism of the ideologies themselves, as neither has the ability to adapt to an environment and a mechanism for optimisation is missing.

Section 4 examined social mobility in the biased interaction game. It revealed that, whilst an emergent hierarchical structure is largely stable, any single agent's position within the hierarchy is far from pre-ordained. Social mobility is an observed feature of many human societies (Tyree et al., 1979), and one might expect any mathematical description of self-organising social structures to allow for it. Far from having to explicitly introduce a mechanism for social mobility into the model, the biased interaction game naturally produces systems with both an emergent hierarchical structure and social mobility within it.

Section 5 looked at the long-term dynamics of the model and demonstrated its non-linear nature. This is an encouraging result, since complex or chaotic system behaviour is a typical observation of real-world systems that is difficult to model explicitly. Inherent non-linearity is a natural property of the biased interaction game.

Finally, section 6 illustrated how biased interaction game-based models might be used to explore real-world scenarios. In this instance, two competing wealth redistribution philosophies were briefly explored (social welfare and a universal basic income). The two models were easy to define, and simple to examine and compare. The results make intuitive sense suggesting that, whilst social welfare addresses inequality and requires a low tax rate, it does so by disrupting the lower and middle bands in the social hierarchy. In moderation, UBI reduces inequality whilst also preserving the underlying structure of the social hierarchy but does so at the cost of a higher tax rate. At extremes of taxation, however, hierarchy collapses, so there are real limits to its applicability. This ability to visualise the system-wide impact of different wealth redistribution approaches may well provide new insight for policy makers.

Simple rules of interaction based on principles of bounded rationality under conditions of scarcity form the foundation of the biased interaction game. This paper's exploration of the game revealed systems with an emergent hierarchical structure characterised by social mobility between bands. That the game can be used to describe such systems is an exciting new result with plenty of scope for further investigation and application.

# Appendix

A number of experiments were conducted to more fully determine the characteristics of the biased interaction game, in particular the emergence of hierarchical structures and agent mobility between the hierarchical bands. To more fully exercise the model, a supercomputer array was employed. Using the supercomputer facility enabled longer experiments with a higher population of agents over a greater number of scenarios. This research utilised Queen Mary's Apocrita HPC facility, supported by QMUL Research-IT. http://doi.org/10.5281/zenodo.438045. We acknowledge the assistance of Giles Greenway of the ITS Research team at Queen Mary University of London.

At the core of these results is the use of the chart described in section 3, plotting the starting vs. ending incumbent values for each agent in the experiment. The results are displayed over a number of pages of charts, with eight charts to each page. One experiment variable is varied across the eight charts in each page of results, and another variable takes on different values for each page. Viewing across a number of consecutive pages allows for a quick interpretation of how the different experimental variables affect the output of the biased interaction game.

Experiment A1:

- $S$ = Chart variable (0.01, 0.05, 0.15, 0.20, 0.30, 0.40, 0.50, 0.60)
- $N$ = Page variable (30, 100, 200, 300, 500, 750)
- $R$ = 0.1
- $I$ = 2,000
- Starting distribution = 'Random'

Figures A1.1 to A1.6 show that the emergent hierarchical structure and the associated Gini coefficient closely follow the scarcity $S$ defined for an experiment and that this property is independent of the number of agents $N$.

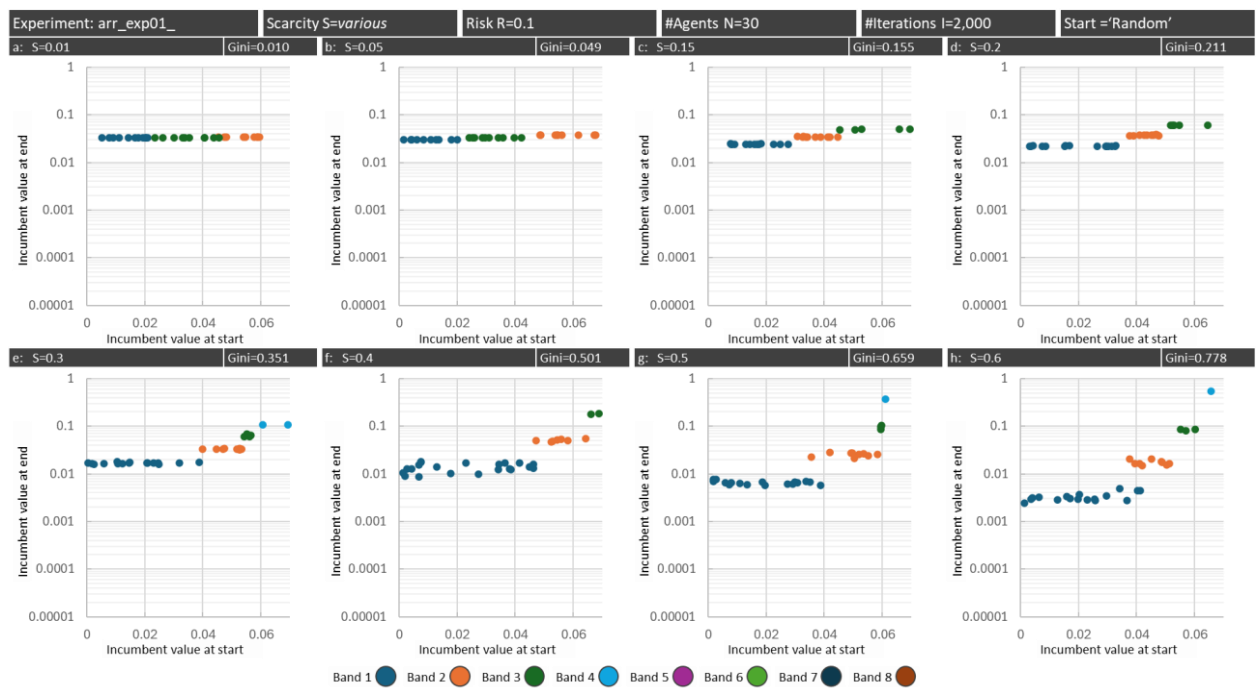

*Figure A1.1 – Page 1: N= 30*

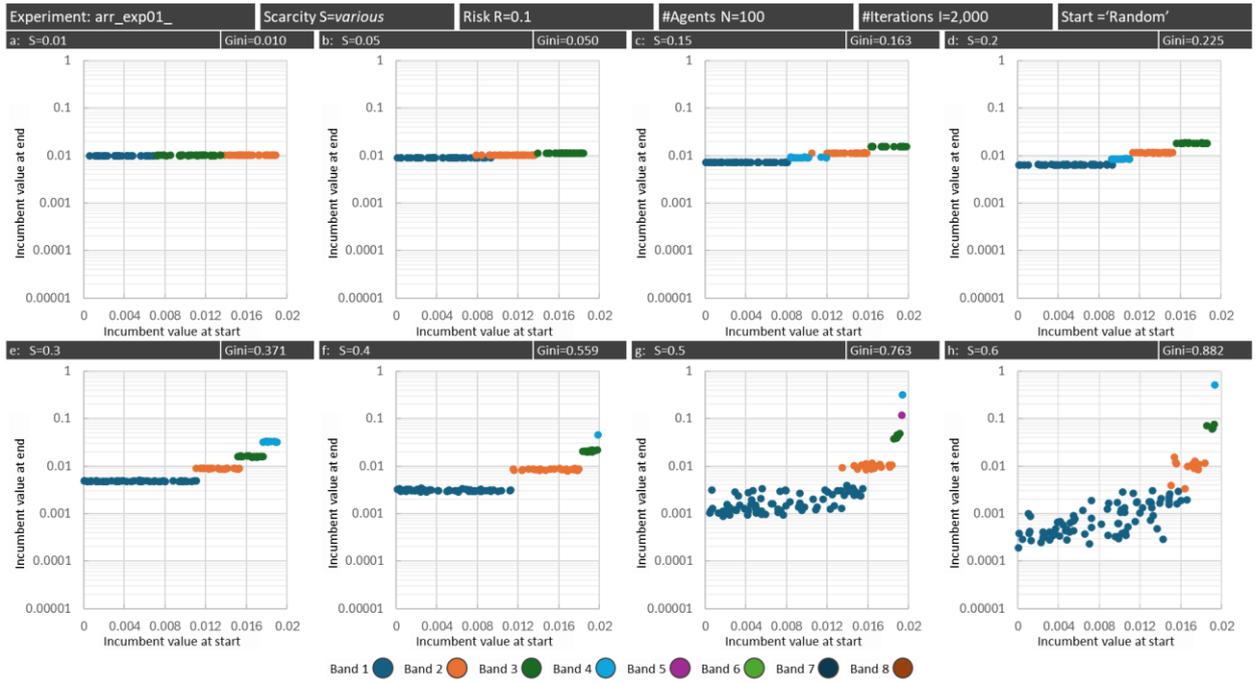

*Figure A1.2 – Page 2: N= 100*

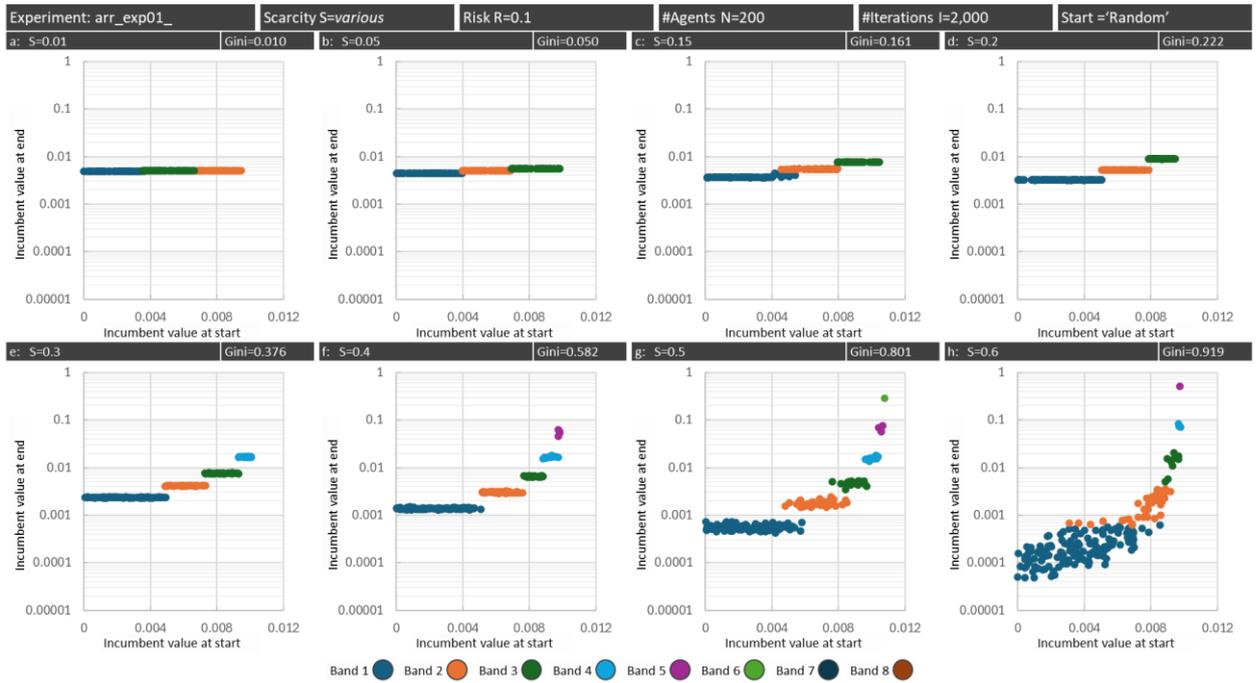

*Figure A1.3 – Page 3: N= 200*

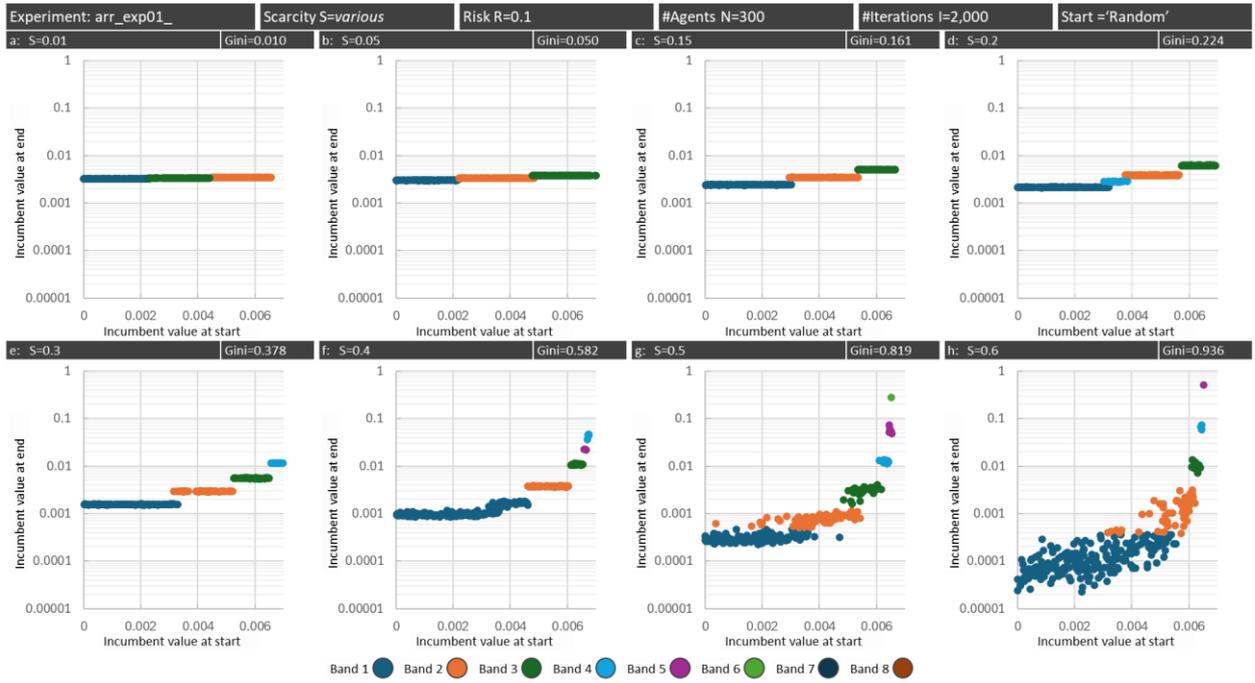

*Figure A1.4 – Page 4: N= 300*

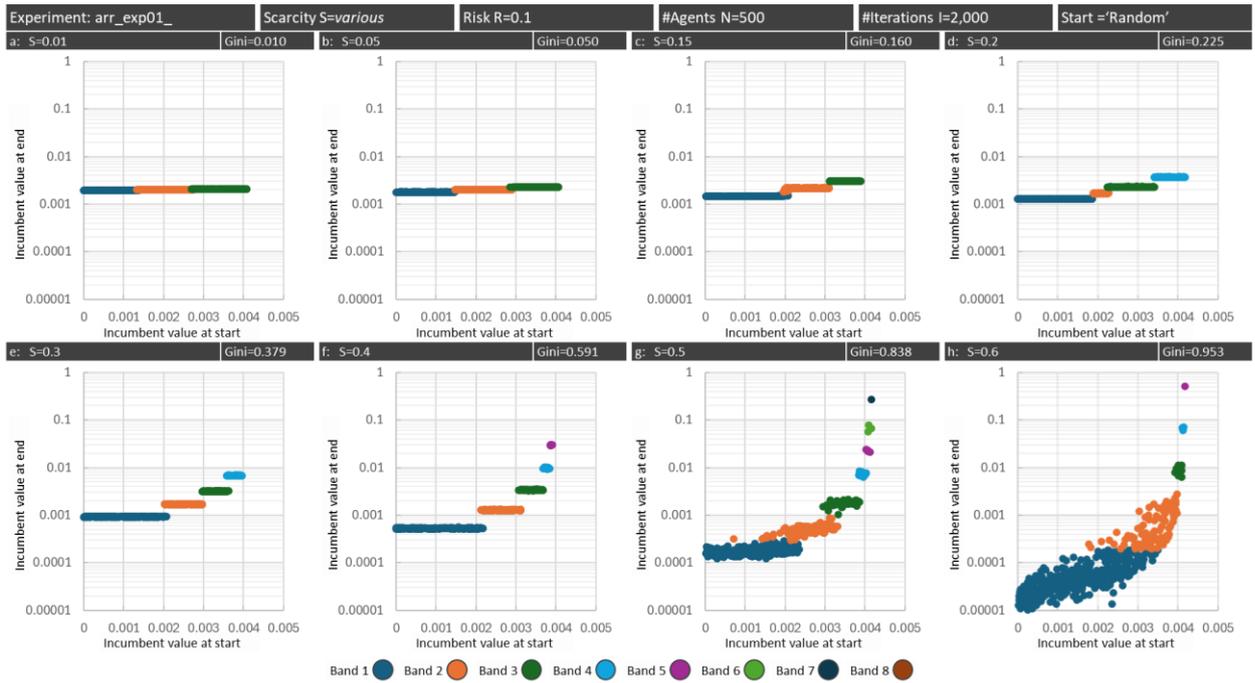

*Figure A1.5 – Page 5: N= 500*

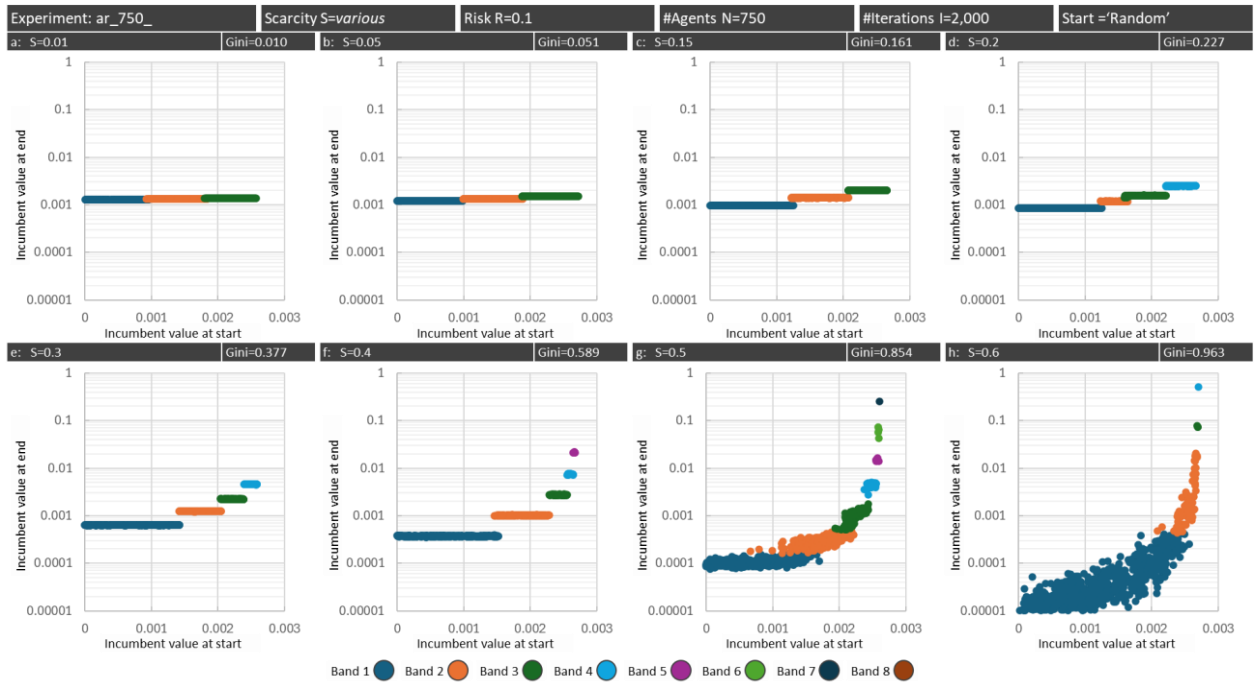

*Figure A1.6 – Page 6: N= 750*

As the number of agents increases, and for higher values of scarcity, it is apparent that the hierarchical banding has yet to fully form. There is however still a clear relationship between the scarcity value and the measured Gini coefficient. It is likely that the number of iterations chosen for this series of experiments was too low to facilitate the emergence of clear hierarchical bands for higher agent populations.

Experiment A2:

- $S$ = Page variable (0.20, 0.30, 0.35, 0.40, 0.45, 0.50, 0.55)
- $N$ = 750
- $R$ = Chart variable (0.05, 0.10, 0.15, 0.20, 0.30, 0.50, 0.70, 0.90)
- $I$ = 2,000
- Starting distribution = 'Random'

Figures A2.1 to A2.7 suggest that, for low levels of scarcity, the emergent hierarchical structure is robust with an agent's ultimate incumbent value being highly dependent on their starting value. This seems unaffected by increasing levels of interaction risk, $R$. As scarcity increases, however, whilst the tendency to band remains, and the emerging degree of inequality measured by the Gini coefficient is consistent for different values of $R$, the influence of an agent's starting incumbent value is reduced. For instance, in Figure A2.5 where $R=0.9$, the two lowest bands are occupied by agents with a wide range of starting incumbent values, with no clear relationship between starting incumbent value and the band ultimately occupied.

In Figure A2.7 where $R=0.9$, whilst the banding has yet to fully develop, the Gini coefficient indicates a degree of inequality consistent with lower levels of $R$ (for the same level of scarcity). However, the ultimate incumbent values for the agents seem completely dissociated from their starting incumbent values. These results suggest that agent mobility across the emergent hierarchical banding structure is increased with increasing levels of scarcity or interaction risk.

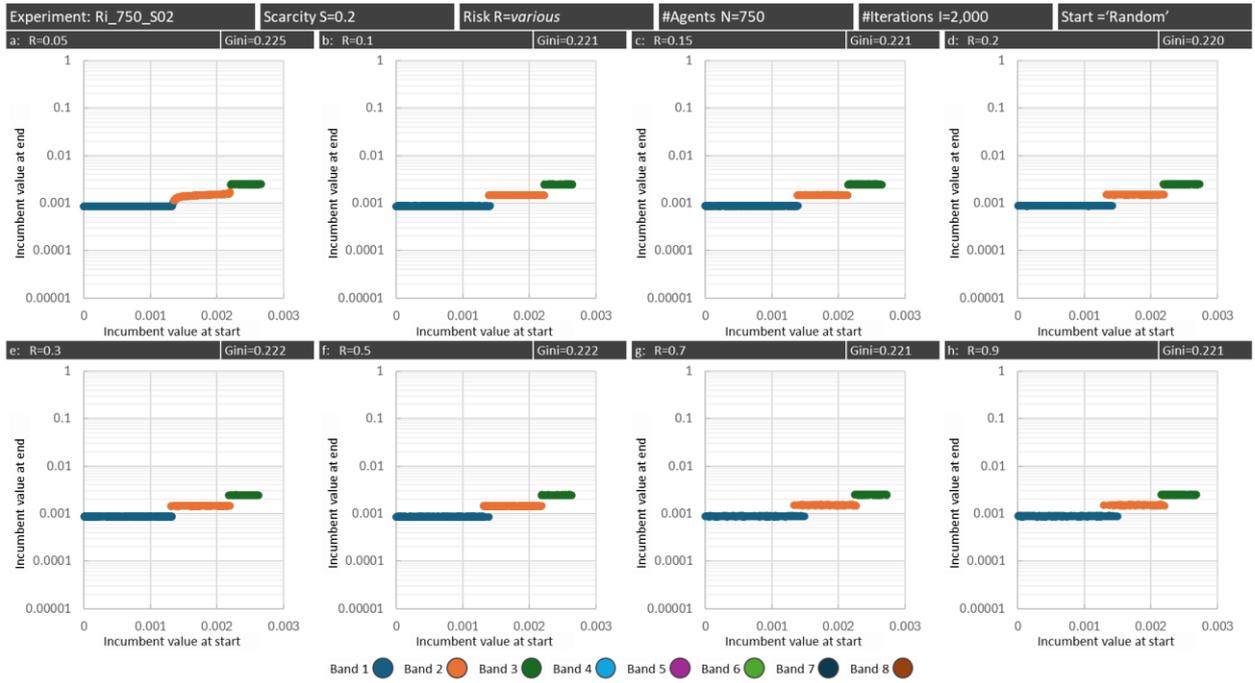

*Figure A2.1 – Page 1: S=0.20*

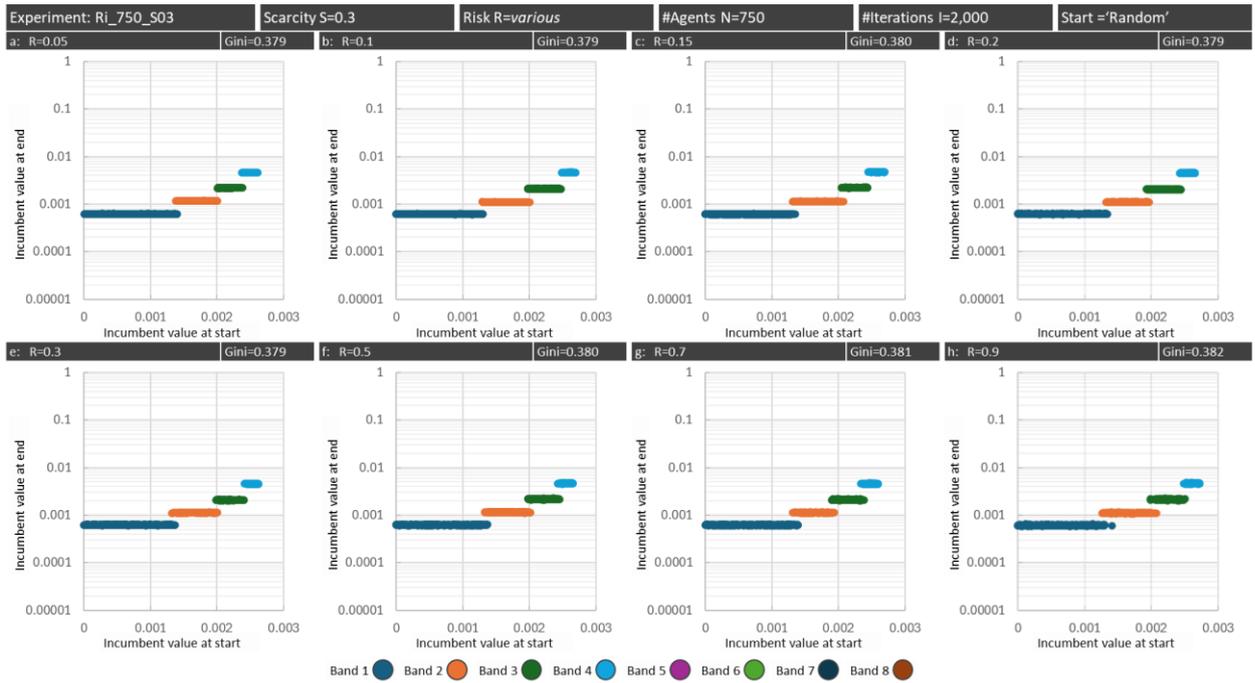

*Figure A2.2 – Page 2: S=0.30*

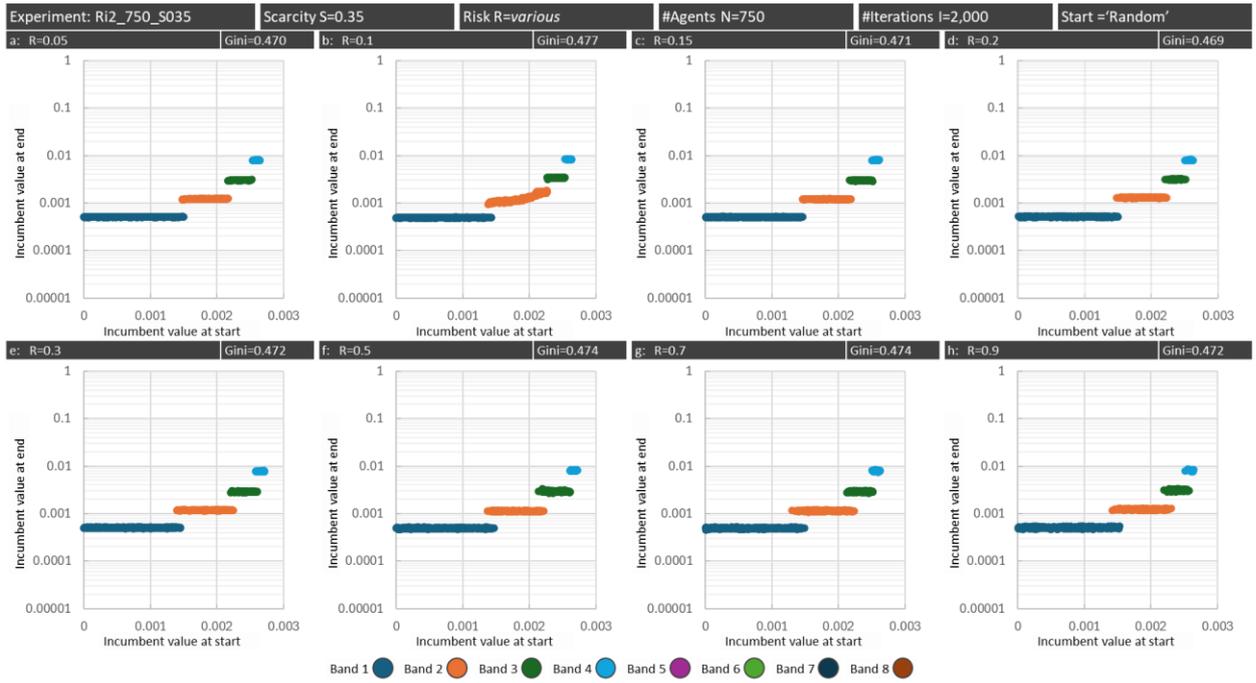

*Figure A2.3 – Page 3: S=0.35*

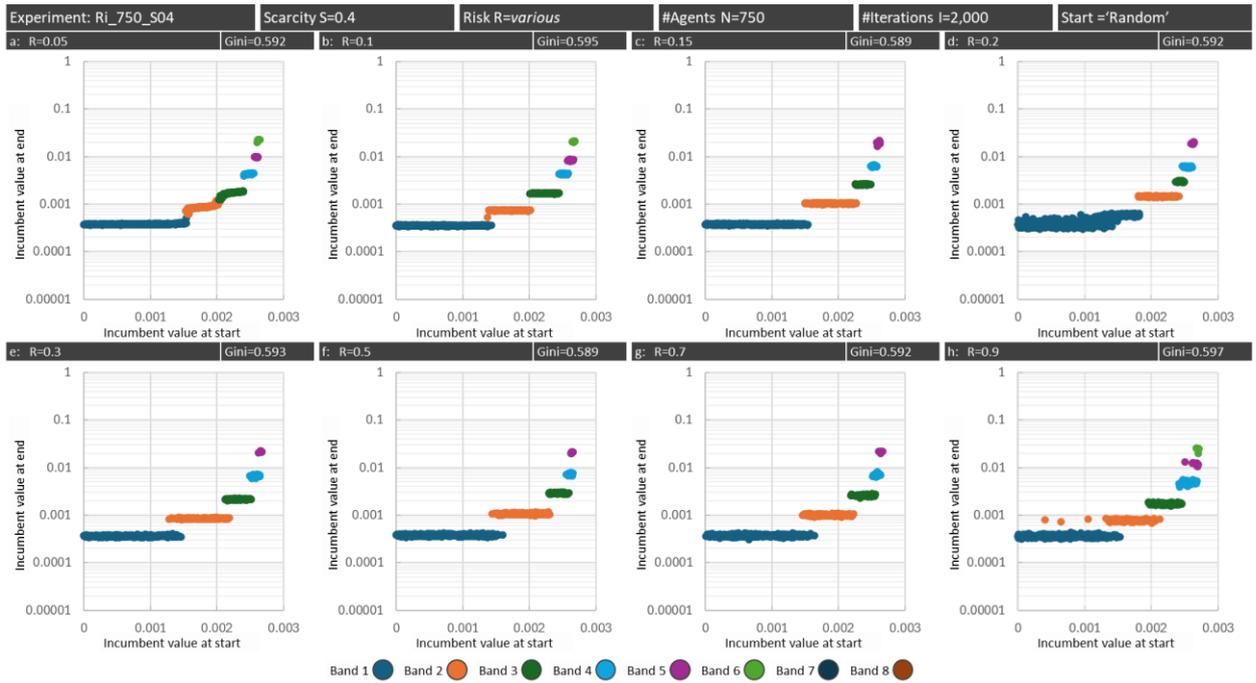

*Figure A2.4 – Page 4: S=0.40*

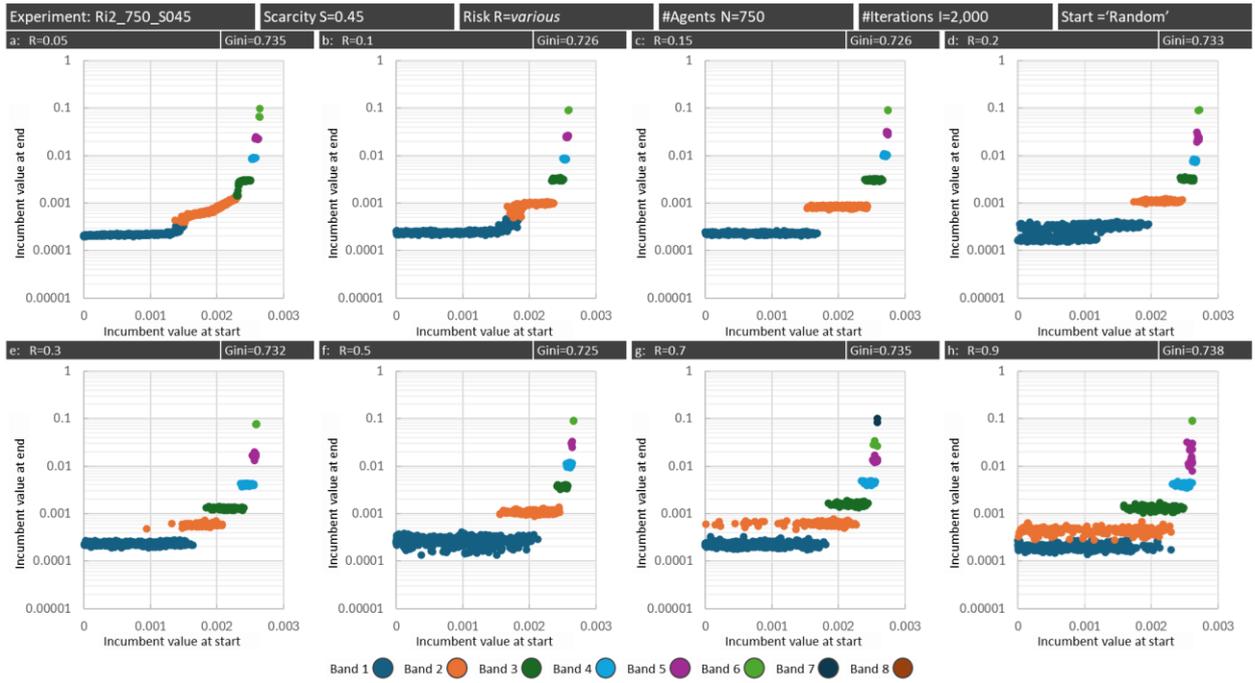

*Figure A2.5 – Page 5: S=0.45*

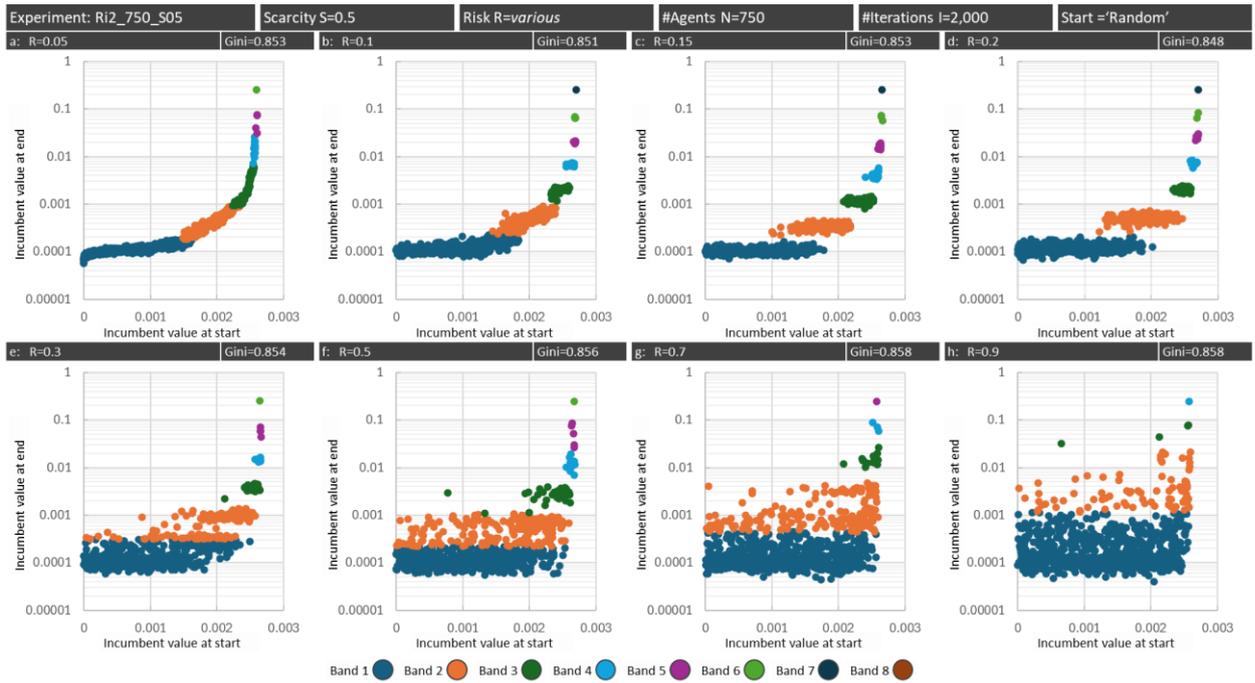

*Figure A2.6 – Page 6: S=0.50*

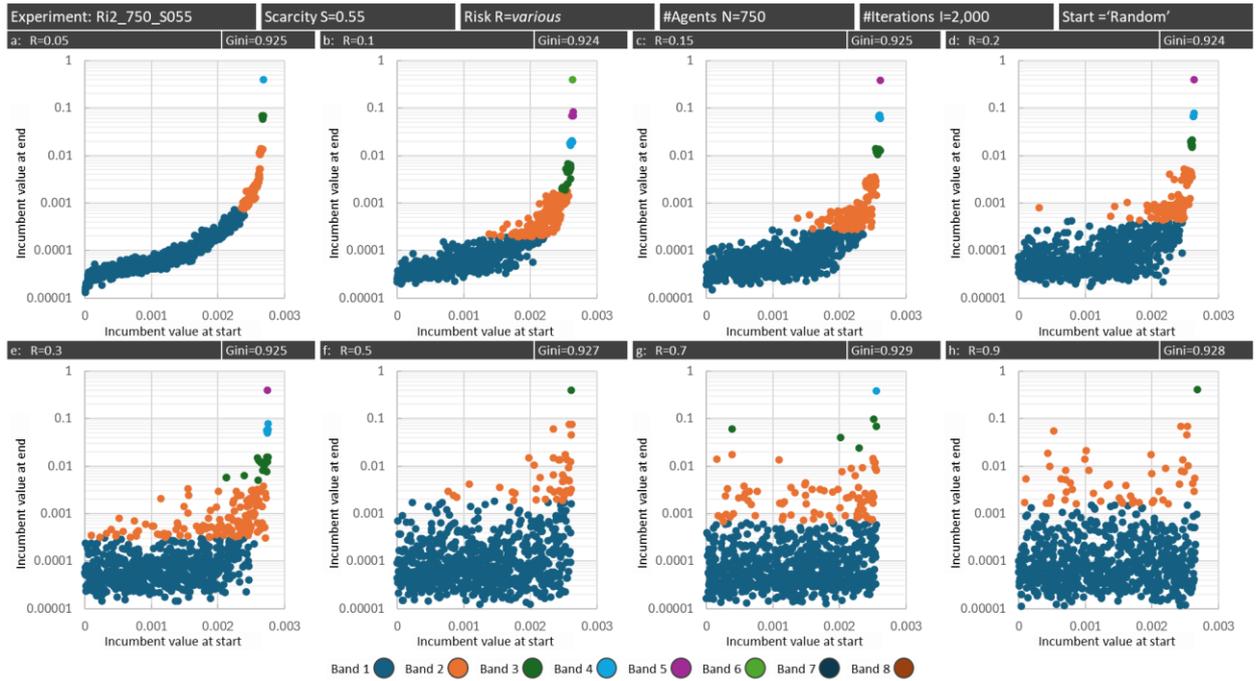

*Figure A2.7 – Page 7: S=0.55*

Experiment A3:

- *S* = Page variable (0.30, 0.40, 0.50)
- *N* = 150
- *R* = Chart variable (0.05, 0.10, 0.15, 0.20, 0.30, 0.50, 0.70, 0.90)
- *I* = 200,000
- Starting distribution = 'Random'

In experiments A1 and A2, the low number of iterations chosen led to the situation where experiments with higher levels of scarcity (or agent population *N*) did not have fully formed hierarchical structures at their conclusion. Experiment A3 has an increased number of experiment iterations (*I=200,000*). The number of agents is reduced to 150 to allow for reasonable experiment run times.

Hierarchical banding is evident in Figures A3.1 to A3.3. Evidence of agent mobility is present, even at the lowest level of scarcity (*S=0.30*) when the interaction risk is high (*R=0.7* and *R=0.9*). Figure A3.2 shows a clear hierarchy when scarcity *S=0.40*, but with a reduction in the influence of agent starting incumbent value, especially with high values of interaction risk. Agent mobility occurs, without necessarily disrupting the emergent hierarchical structure. When scarcity increases (Figure A3.3) the hierarchical bands that emerge are far fuzzier: there is more variability in the incumbent values of the agents within each band. Since this result is occurring after 200,000 iterations of the model, this can reasonably be thought of as a steady state. Again, the influence of an agent's starting incumbent value appears greater for lower values of interaction risk but reduces as the interaction risk increases.

One concludes that, in the steady state, the biased interaction game leads to systems with an emergent hierarchical structure whose depth is dependent on the level of scarcity in the environment, but that still allows for agent mobility between hierarchical bands with a likelihood that increases with increasing scarcity and interaction risk.

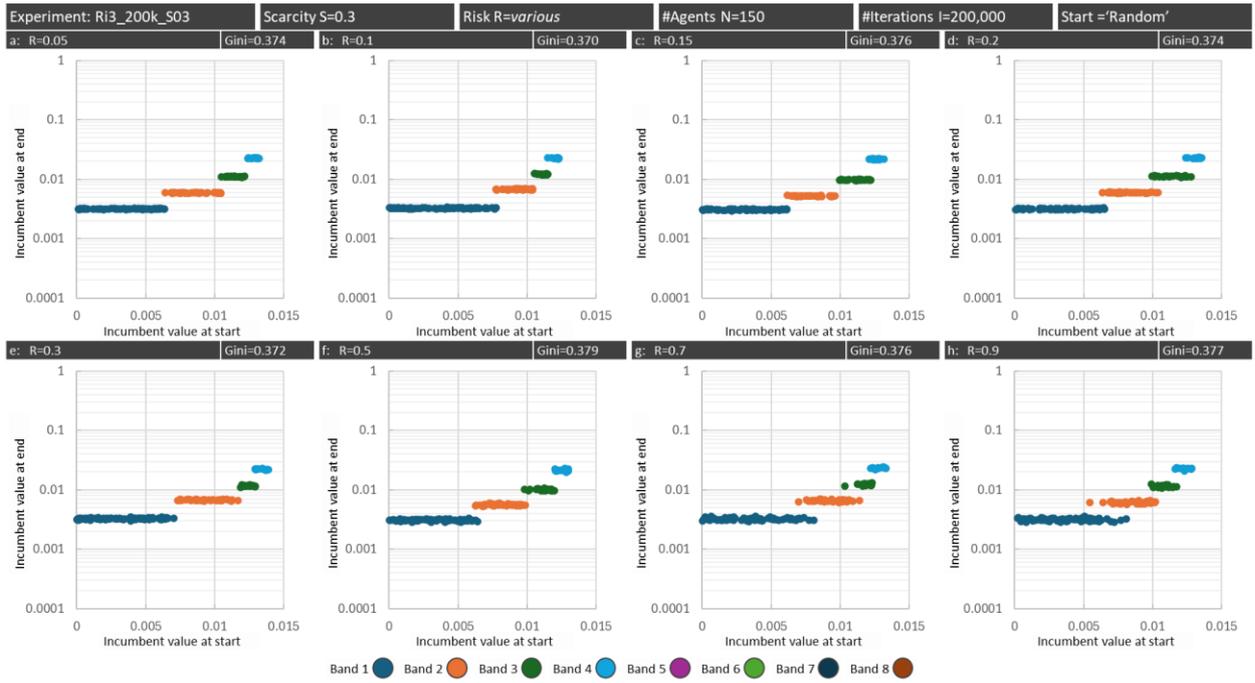

*Figure A3.1 – Page 1: S=0.30*

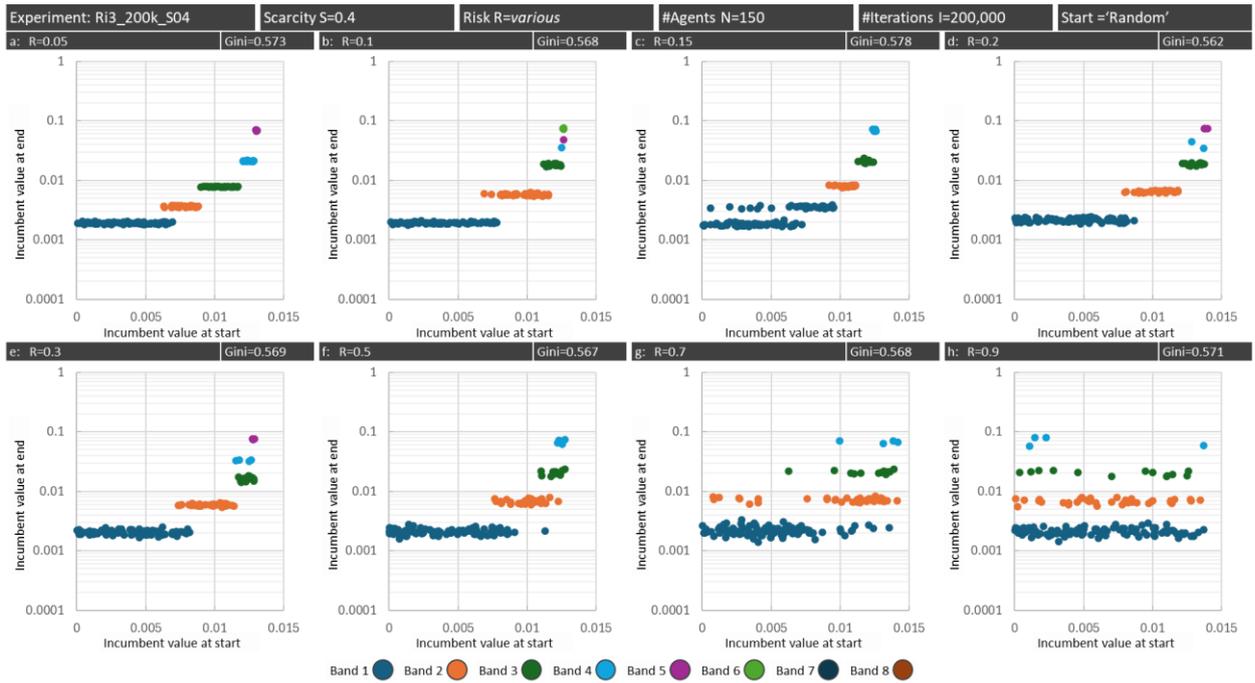

*Figure A3.2 – Page 2: S=0.40*

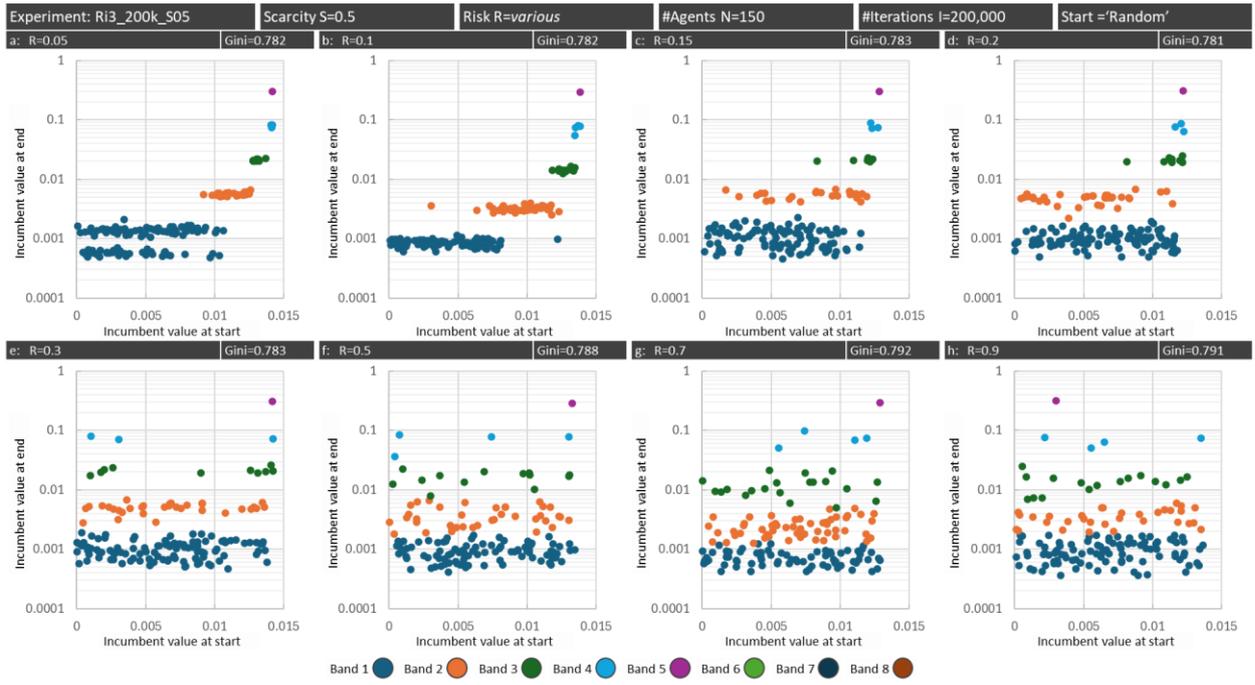

*Figure A3.3 – Page 3: S=0.50*